\documentclass[aps,pra,preprint,longbibliography]{revtex4-1} 

\usepackage[centertags]{amsmath}
\usepackage{graphicx}
\usepackage{amsfonts}
\usepackage{epstopdf}
\usepackage{color}  

\usepackage{subcaption} 
\usepackage[export]{adjustbox} 
\usepackage{cleveref}

\begin{document}

\title{Directional Soliton and Breather Beams} 
\author{A. Chabchoub$^{1,2\ast}$, K. Mozumi$^{2}$, N. Hoffmann$^{3,4}$, A. V. Babanin$^{5}$, A. Toffoli$^{5}$, J. N. Steer$^{6}$, T. S. van den Bremer$^{7}$, N. Akhmediev$^{8}$, M. Onorato$^{9}$ and T. Waseda$^{2}$} 
\affiliation{$^1$ Centre for Wind, Waves and Water, School of Civil Engineering, The University of Sydney, Sydney, NSW 2006, Australia}
\email{amin.chabchoub@sydney.edu.au}
\affiliation{$^2$ Department of Ocean Technology Policy and Environment, Graduate School of Frontier Sciences, The University of Tokyo, Kashiwa, Chiba 277-8563, Japan}
\affiliation{$^3$ Dynamics Group, Hamburg University of Technology, 21073 Hamburg, Germany}
\affiliation{$^4$ Department of Mechanical Engineering, Imperial College London, London SW7 2AZ, United Kingdom}
\affiliation{$^5$ Department of Infrastructure Engineering, The University of Melbourne, Parkville, VIC 3010, Australia}
\affiliation{$^6$ School of Engineering, University of Edinburgh, Edinburgh EH9 3FB, United Kingdom}
\affiliation{$^7$ Department of Engineering Science, University of Oxford, Oxford OX1 3PJ, United Kingdom}
\affiliation{$^8$ Research School of Physics and Engineering, The Australian National University, Canberra, ACT 2000, Australia}
\affiliation{$^9$ Dipartimento di Fisica, Universit\`a degli Studi di Torino and INFN, 10125 Torino, Italy}

\begin{abstract} 
Solitons and breathers are nonlinear modes that exist in a wide range of physical systems. They are fundamental solutions of a number of nonlinear wave evolution equations, including the uni-directional nonlinear Schr\"odinger equation (NLSE). We report the observation of slanted solitons and breathers propagating at an angle with respect to the direction of propagation of the wave field. As the coherence is diagonal, the scale in the crest direction becomes finite, consequently, a beam dynamics forms. Spatio-temporal measurements of the water surface elevation are obtained by stereo-reconstructing the positions of the floating markers placed on a regular lattice and recorded with two synchronized high-speed cameras. Experimental results, based on the predictions obtained from the (2D+1) hyperbolic NLSE equation, are in excellent agreement with the theory. Our study proves the existence of such unique and coherent wave packets and has serious implications for practical applications in optical sciences and physical oceanography. Moreover, unstable wave fields in this geometry may explain the formation of directional large amplitude rogue waves with a finite crest length within a wide range of nonlinear dispersive media, such as Bose-Einstein condensates, plasma, hydrodynamics and optics.
\end{abstract}
\maketitle 
\section*{Introduction}
Ocean waves are complex two-dimensional dynamical structures that cannot be easily modelled in their full complexity. Variations of depth, wind strength, wave breaking, randomness and large amplitude waves add tremendously to this complexity \cite{komen1996dynamics}. Despite these complications, the research on water waves is important and significant advances have been made so far \cite{rintoul2018global}. The progress is mainly due to simplified models that are used to analyse their dynamics \cite{osborne1980internal}. Moreover, validity of these models can be confirmed in down-scaled experiments in water wave facilities that do exist in many research laboratories around the world. These experiments are crucially significant to build our understanding of larger scaled oceanic waves. Evolution equations and their solutions are essential for water wave modeling, while computerized equipment is a key for their accurate generation.

One of the essential complications in ocean wave dynamics is the unavoidable existence of two horizontal spatial coordinates. Directional behaviors of the surface waves in nature are of principal importance for practical applications ranging from wave forecast through modeling air-sea interactions and to, most importantly, environmental and optical sciences. In a simplified way, such wave field consists of many waves crossing each other at various angles implying at a linear level that the water surface is a mere interference of short- and long-crested waves coming from different directions \cite{onorato2009statistical,toffoli2011extreme,pinho2015emergence}. Here, we leave aside these complexities. Instead, we start with a simple question: what does the second coordinate add to the dynamics when the waves are mostly uni-directional? This simple question must be answered before considering more complicated cases. 

Indeed, uni-directional nonlinear wave dynamics on the water surface in deep-water, that is, assuming that the water depth is significantly larger than the waves' wavelength, can be described by the nonlinear Schr\"odinger equation (NLSE) that takes into account dispersion and nonlinearity \cite{zakharov1968stability}. Being an integrable evolution equation, it allows for the study of particular and localized coherent wave patterns, such as solitons and breathers \cite{akhmediev1997solitons,osborne2010nonlinear,blanco2016pure}. The latter are of major relevance to study the fundamental wave dynamics in nonlinear dispersive media with a wide range of applications \cite{solli2007optical,onorato2013roguereport,dudley2014instabilities}. While the NLSE has been formulated for planar waves and wave packets propagating in the same direction as the underlying carrier waves, there is also a generalization of the framework, the so-called directional NLSE, which allows the envelope and homogeneous planar carrier wave to propagate at an angle to each other. This possibility adds novel and unexpected features to well-known nonlinear and coherent wave propagation motions as we examine in this work. Unfortunately, from the theoretical perspective, the directional deep-water NLSE is not integrable. As a consequence, these nontrivial nonlinear solutions are not easy to identify. Early attempts to generate some nonlinear states were based on symmetry considerations \cite{kartashov2011solitons}. It has been shown \cite{saffman1978stability,yuen1982nonlinear} that each uni-directional solution of the NLSE has a family counterpart solutions for which the packet beam propagates obliquely to the short-crested carrier wave.

These type of wave processes are directly relevant in oceanography \cite{longuet1976nonlinear,toffoli2010development,borge2013detection}. However, taking into account many areas in physics for which the NLSE is the fundamental governing equation, our ideas can be bluntly expanded to these fields such as Bose-Einstein condensates, plasma and optics \cite{nguyen2017formation,cabrera2018quantum,cheiney2018bright,semeghini2018self,kartashov2018three}.

In the present study, we report an experimental framework and observations of hydrodynamic diagonal solitons and breathers in a deep-water wave basin. Our results confirm and prove the existence of such unique and coherent beams of quasi-one-dimensional and short-crested wave group in a nonlinear dispersive medium. 
\section*{Results}

\begin{figure}[tb]
\centering
\begin{subfigure}[t]{0.01\columnwidth}\textbf{a}
\end{subfigure}
\begin{subfigure}[t]{0.47\columnwidth}
  \includegraphics[width=0.9\columnwidth,valign=t]{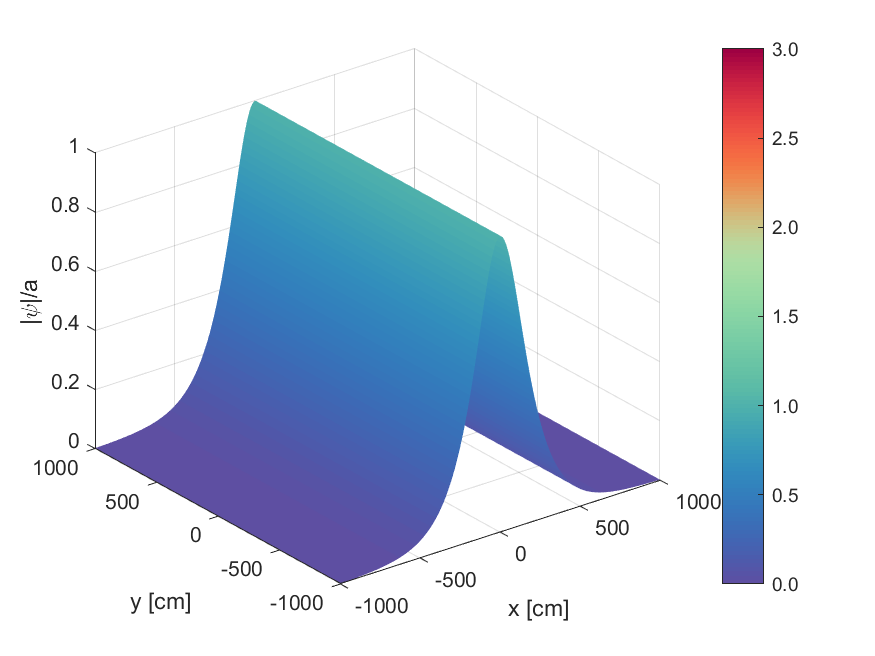} 
\end{subfigure}
\begin{subfigure}[t]{0.01\columnwidth}\textbf{b}
\end{subfigure}
\begin{subfigure}[t]{0.47\columnwidth}
	\includegraphics[width=0.9\columnwidth,valign=t]{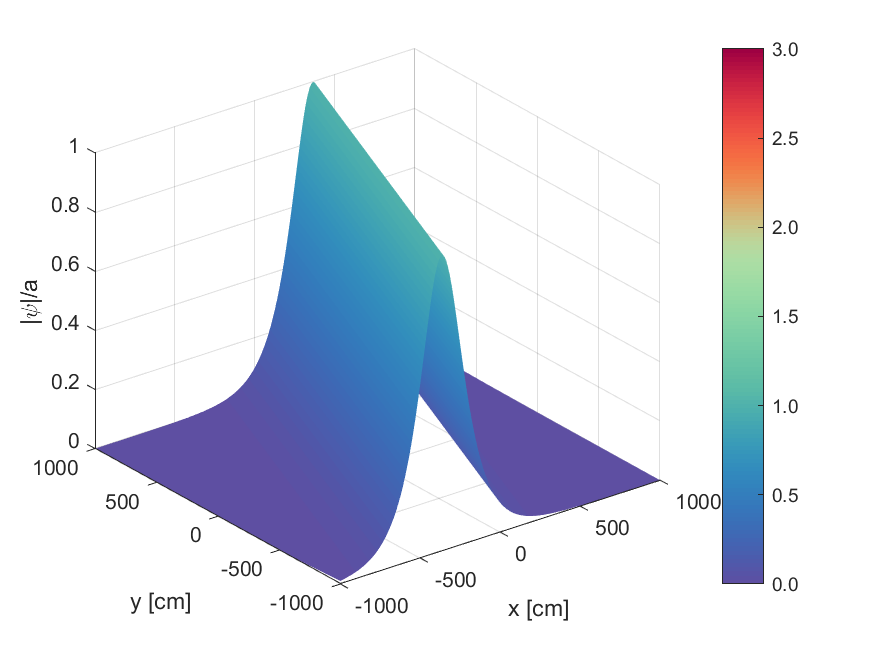}
\end{subfigure}
\begin{subfigure}[t]{0.015\columnwidth}\textbf{c}
\end{subfigure}
\begin{subfigure}[t]{0.47\columnwidth}
  \includegraphics[width=0.9\columnwidth,valign=t]{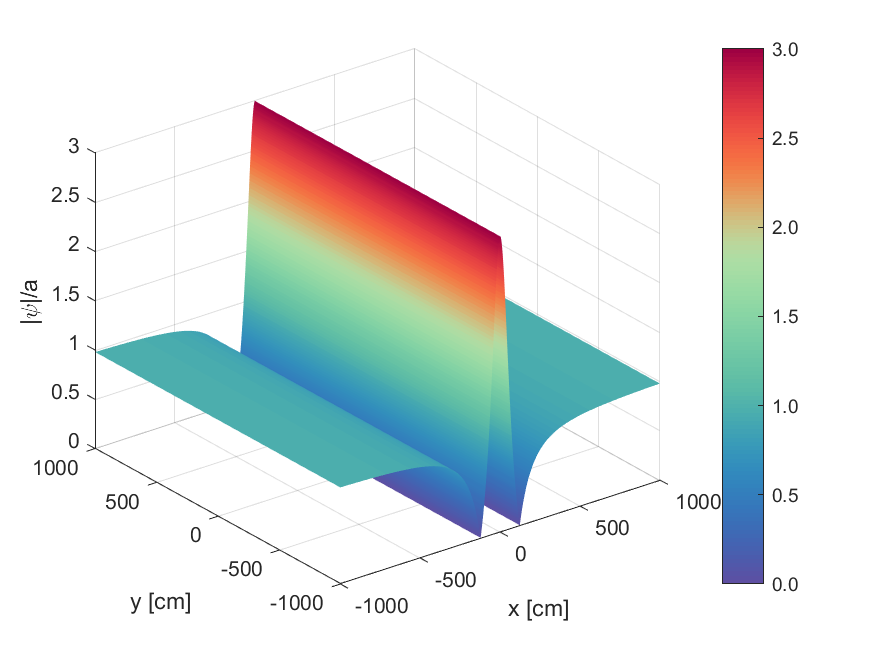} 
\end{subfigure}
\begin{subfigure}[t]{0.01\columnwidth}\textbf{d}
\end{subfigure}
\begin{subfigure}[t]{0.47\columnwidth}
	\includegraphics[width=0.9\columnwidth,valign=t]{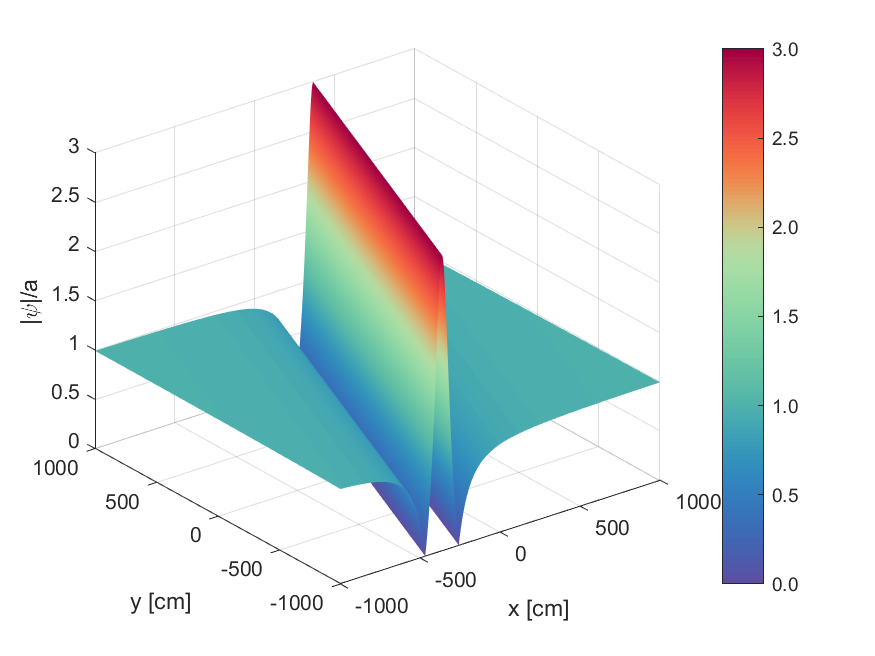}
\end{subfigure}
\caption{Normalized representation of a uni-directional as well as slanted NLSE envelope soliton and Peregrine breather for a carrier amplitude $a=0.02$ m and steepness $ak=0.1$ at $t=0$. {\bf a} Uni-directional envelope soliton dynamics for $\vartheta=0^\circ$. {\bf b} Envelope soliton dynamics slanted by an angle of $\vartheta=20^\circ$. {\bf c} Uni-directional Peregrine breather dynamics for $\vartheta=0^\circ$. {\bf d} Peregrine breather dynamics slanted by an angle of $\vartheta=20^\circ$.}
\label{fig1}
\end{figure}

In order to illustrate such type of universal and directional wave packet, in Fig. \ref{fig1} we show an example of the dimensional shape of an envelope soliton and a Peregrine breather, as parametrized in \cite{shabat1972exact,peregrine1983water}, with amplitude $a=0.02$ m propagating at zero diagonal angle Fig. \ref{fig1}{\bf a} and {\bf c} and at an angle of $\vartheta=20^\circ$ Fig. \ref{fig1}{\bf b} and {\bf d} with respect to the carrier wave whose steepness is $ak=0.1$, see Methods for theoretical framework details.  
% Figure 1

%Figure 2
\begin{figure}[h]
\centering
  \includegraphics[width=0.8\columnwidth]{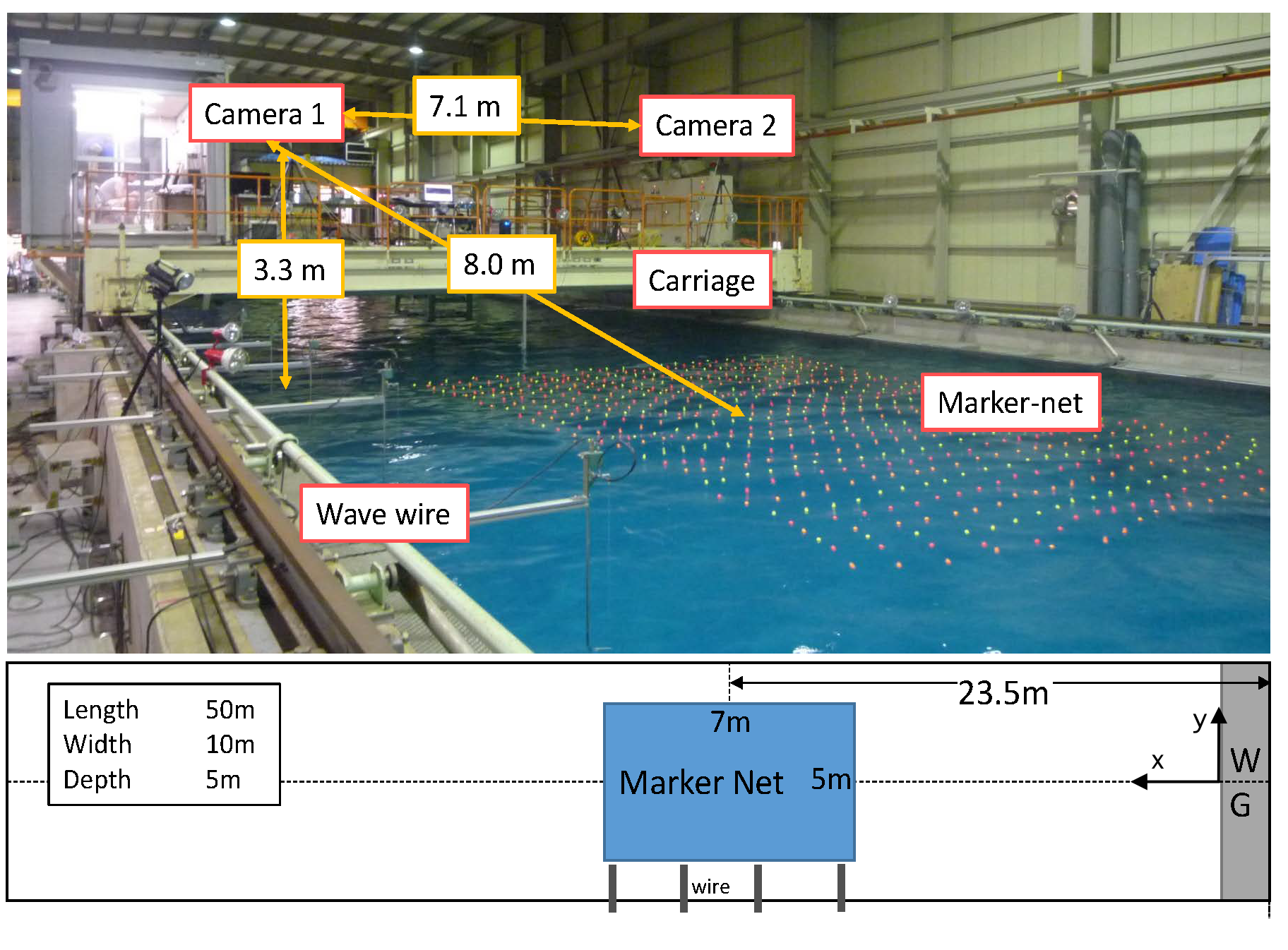} 
  \caption{Experimental set-up. The picture and the sketch show the dimensions of the flume, the location of marker's grid and positions of the two stereo cameras.}
\label{fig2}
\end{figure}

The experiments have been performed in a directional wave basin, installed at the University of Tokyo. Its dimensions are $50\times 10\times 5$ m$^3$. In order to measure the directional wave evolution, a marker-net has been deployed at the center of the basin. The motion of the markers was recorded by two high-speed cameras with a resolution of $2048\times1080$ pixels at 100 frames per second. The two cameras are fully synchronized and are separated by 7.1 m distance across the tank in $y$-direction and positioned at 3.3 m from mean water level and about 10 m away from the center of the marker net in $x$-direction. Moreover, a series of wave wires were installed along the basin in order to follow the wave dynamics along $x$ co-ordinate. These have been placed at 5.21 m, 9.20 m, 10.97 m, 14.01 m, 17.16 m, 20.15 m, 23.02 m, 27.04 m, 28.91 m and 32.05 m from the directional plunger-type wave maker, which consists of 32 sections. Each plunger has a width of 32 cm. More details on the methodology adopted for the data acquisition can be found in \cite{mozumi20153d}. A picture and a sketch of the experimental set-up and the co-ordinate system adopted are depicted in Fig. \ref{fig2}. 

We emphasize that due to the significant size of the digitally collected data, the stereo-reconstruction, that includes an interpolation process, is very challenging \cite{mozumi20153d}. 
 
% Figure 3
\begin{figure}[tb]
\centering
\begin{subfigure}[t]{0.01\columnwidth}\textbf{a}
\end{subfigure}
\begin{subfigure}[t]{0.47\columnwidth}
  \includegraphics[width=0.49\columnwidth,valign=t]{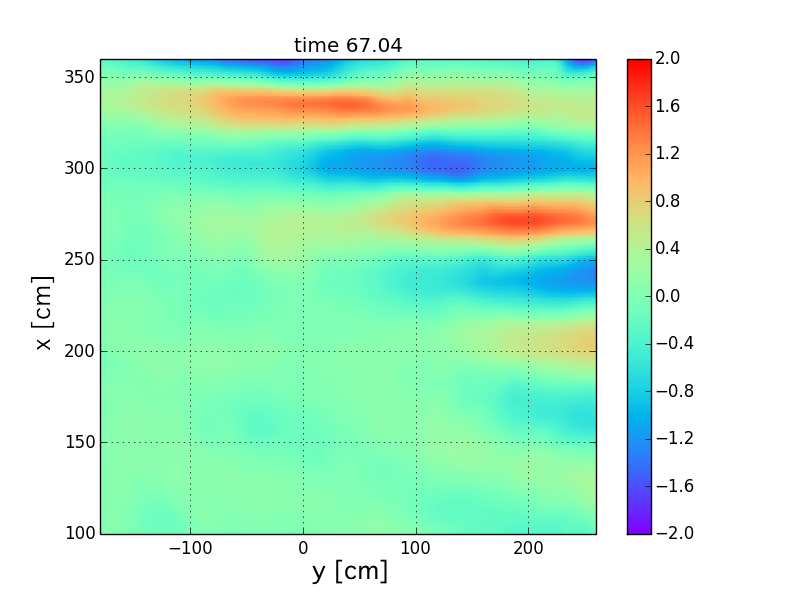} 
	\includegraphics[width=0.49\columnwidth,valign=t]{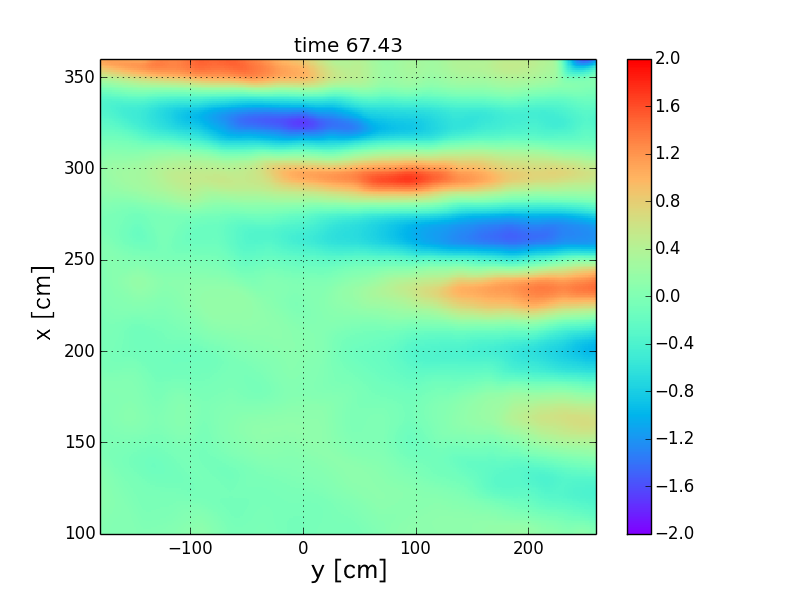}\\
	\includegraphics[width=0.49\columnwidth,valign=t]{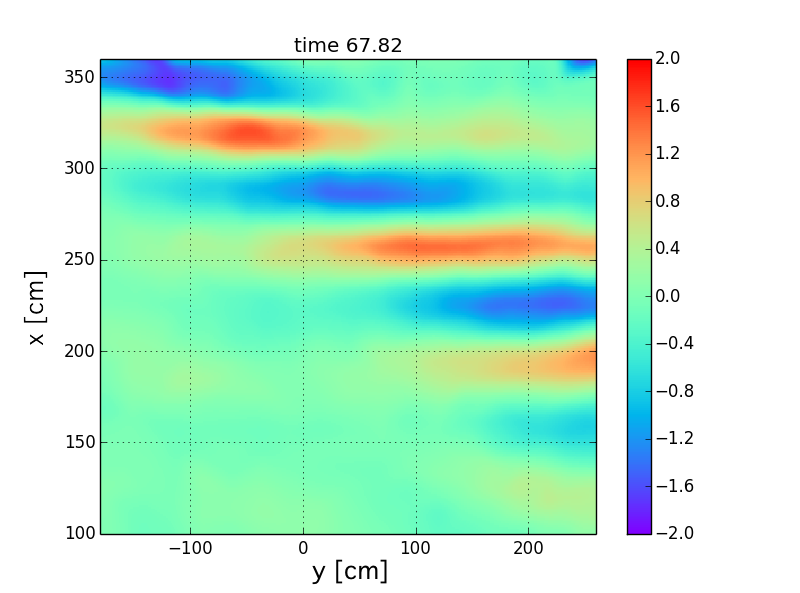} 
	\includegraphics[width=0.49\columnwidth,valign=t]{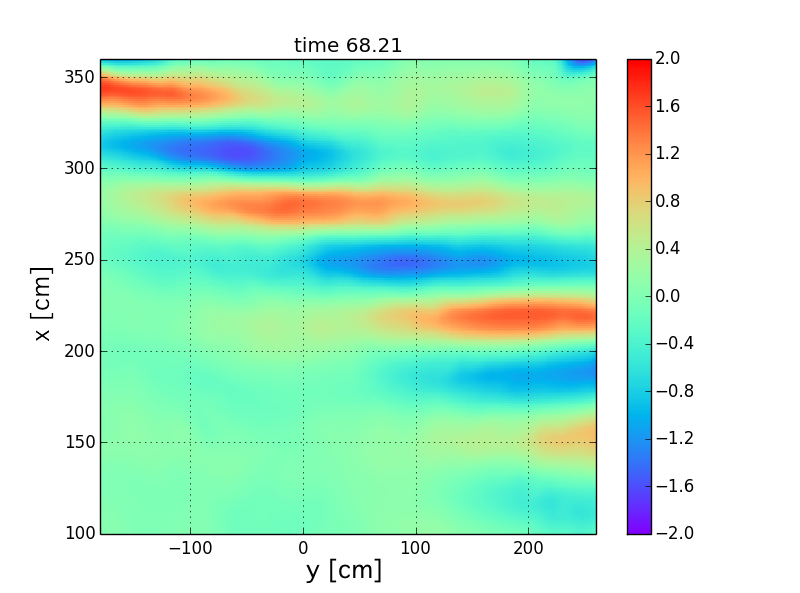}\\
	\includegraphics[width=0.49\columnwidth,valign=t]{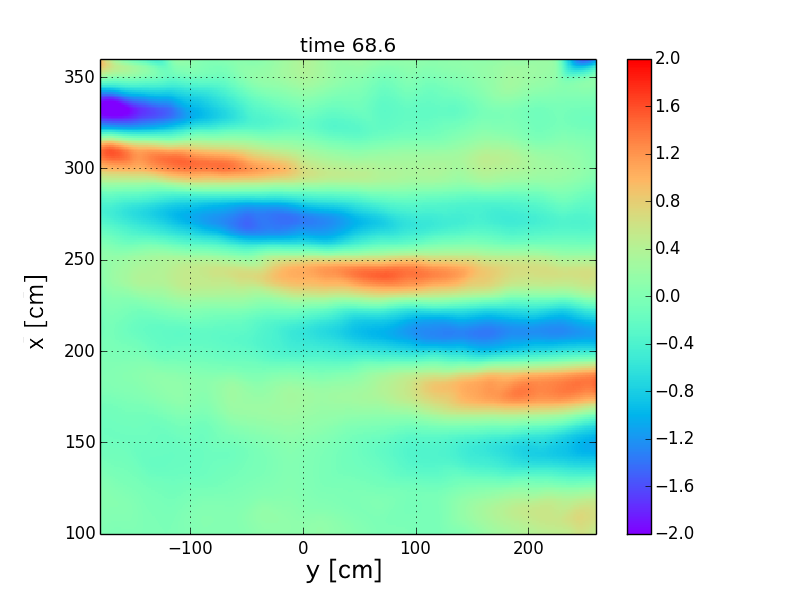} 
	\includegraphics[width=0.49\columnwidth,valign=t]{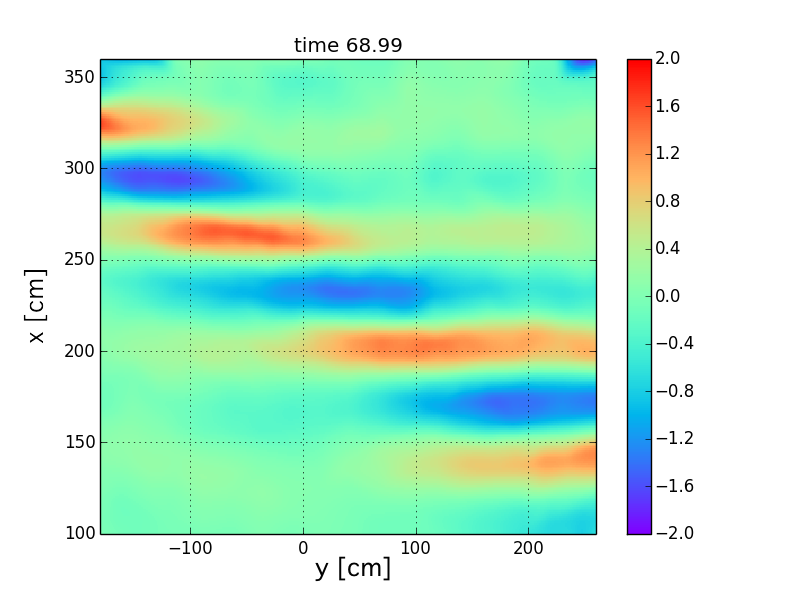}
	\end{subfigure}	
\begin{subfigure}[t]{0.01\columnwidth}\textbf{b}
\end{subfigure}
\begin{subfigure}[t]{0.47\columnwidth}
  \includegraphics[width=0.49\columnwidth,valign=t]{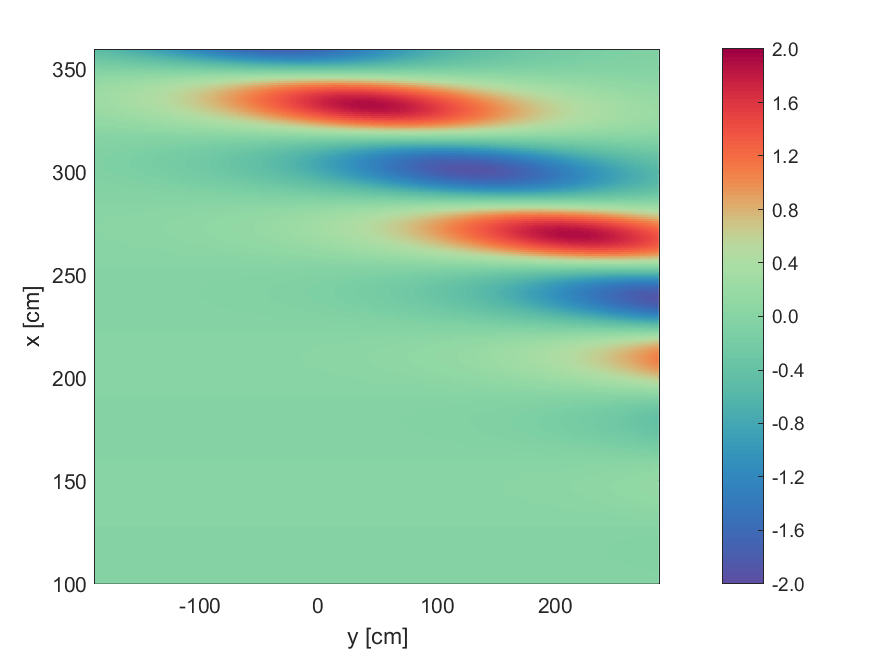} 
	\includegraphics[width=0.49\columnwidth,valign=t]{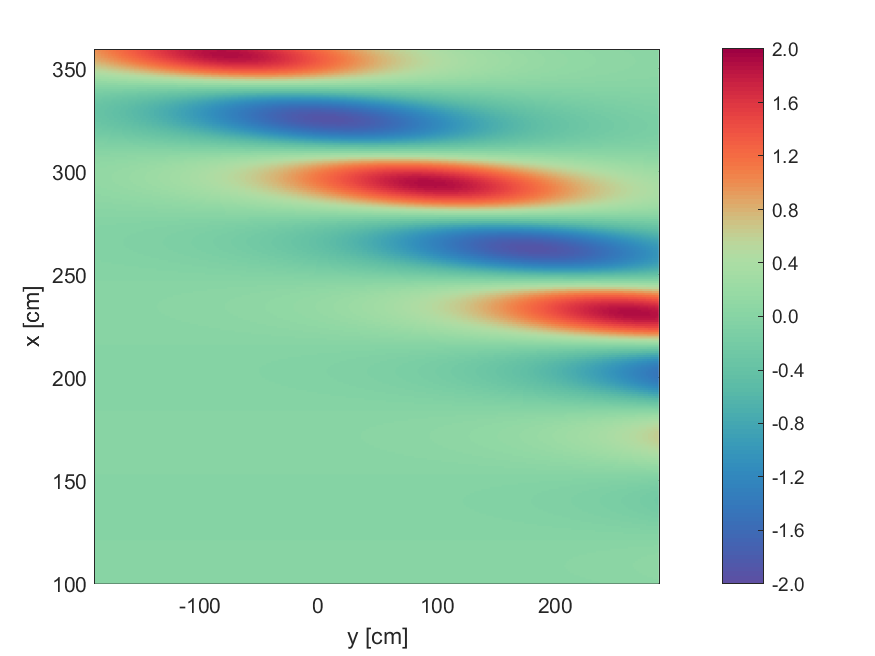}\\
	\includegraphics[width=0.49\columnwidth,valign=t]{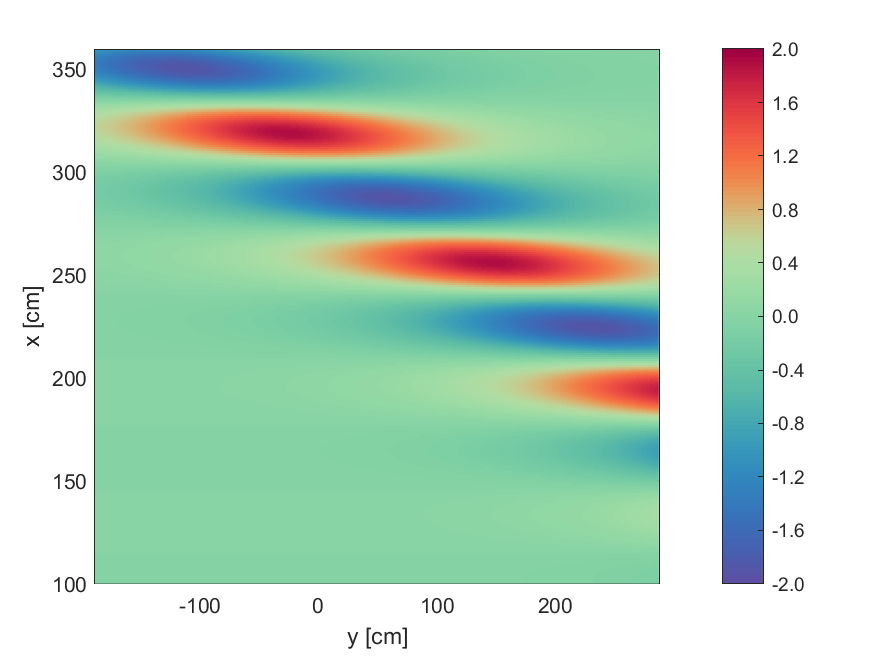} 
	\includegraphics[width=0.49\columnwidth,valign=t]{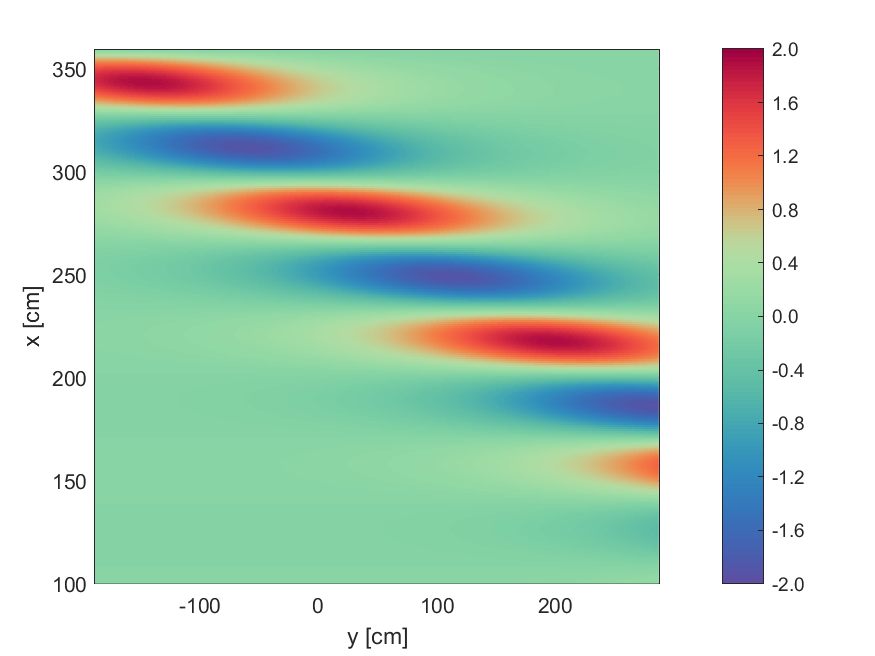}\\
	\includegraphics[width=0.49\columnwidth,valign=t]{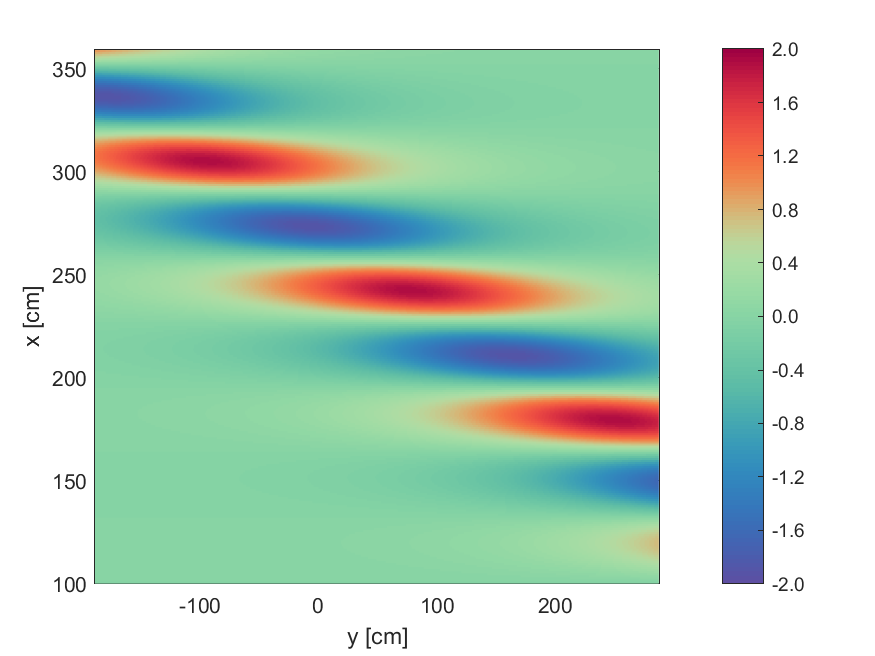} 
	\includegraphics[width=0.49\columnwidth,valign=t]{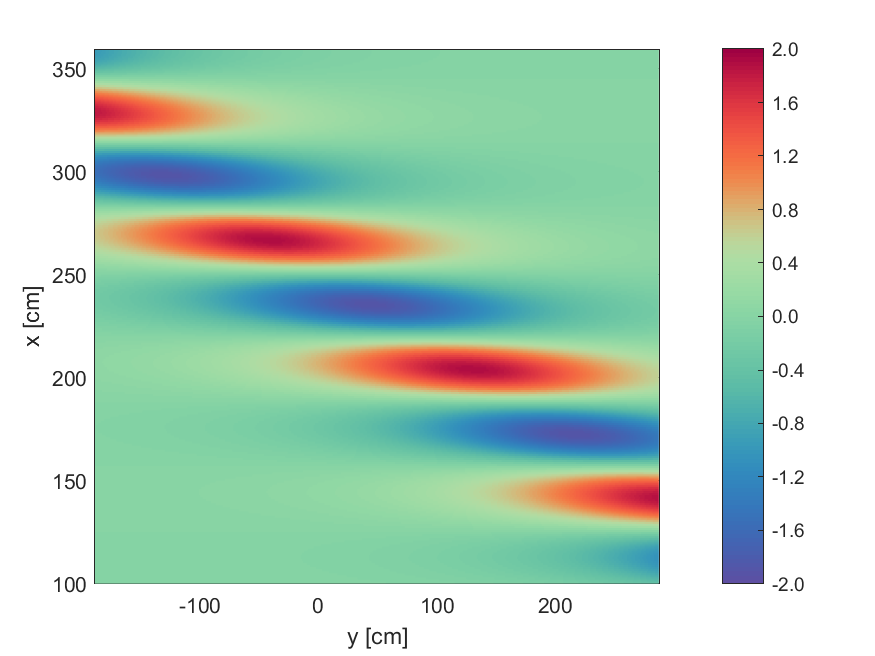}
\end{subfigure}
\caption{
Evolution of a slanted envelope soliton propagating obliquely relative to the carrier wave. The amplitude, expressed in cm, is represented in color scale and the snapshots of the surface elevation are shown at intervals of time $\Delta t=0.39$ s. The parameters adopted are $a=0.02$ m, $ak=0.2$ and $\vartheta=20^\circ$. {\bf a} Stereo-reconstructed surface elevation of the deep-water soliton, propagating in the wave basin. {\bf b} Analytical solution of the corresponding NLSE surface elevation of the slanted coherent structure at the same time intervals. 
}
\label{fig3}
\end{figure} 
%Figure 4
\begin{figure}[tb] 
\centering
\begin{subfigure}[t]{0.015\columnwidth}\textbf{a}
\end{subfigure}
\begin{subfigure}[t]{0.4\columnwidth}
  \includegraphics[width=0.98\columnwidth,valign=t]{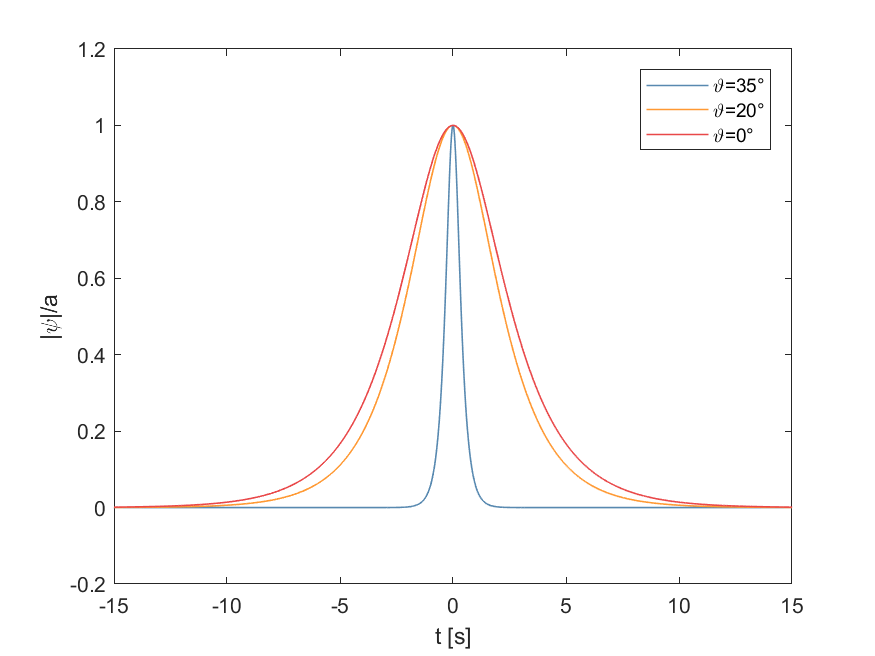} 
\end{subfigure}
\begin{subfigure}[t]{0.015\columnwidth}\textbf{b}
\end{subfigure}
\begin{subfigure}[t]{0.4\columnwidth}
	\includegraphics[width=0.98\columnwidth,valign=t]{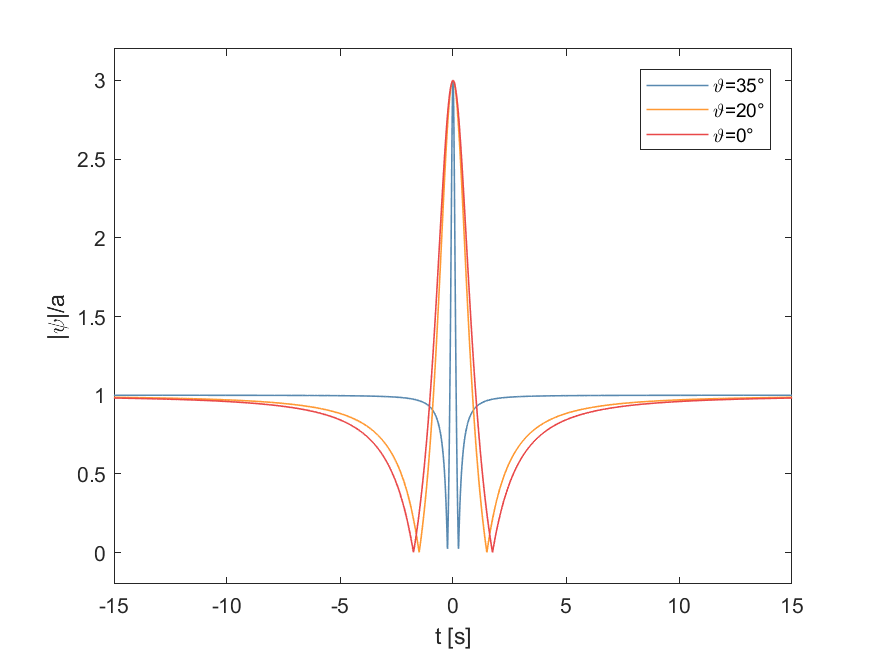}
\end{subfigure}\\
\begin{subfigure}[t]{0.015\columnwidth}\textbf{c}
\end{subfigure}
\begin{subfigure}[t]{0.4\columnwidth}
  \includegraphics[width=0.98\columnwidth,valign=t]{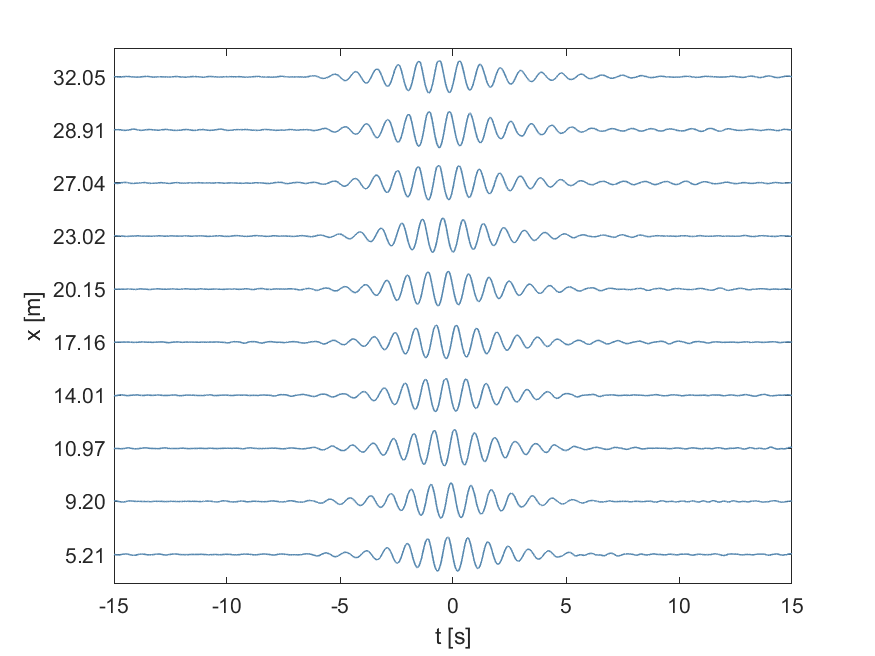} 
\end{subfigure}
\begin{subfigure}[t]{0.015\columnwidth}\textbf{d}
\end{subfigure}
\begin{subfigure}[t]{0.4\columnwidth}
	\includegraphics[width=0.98\columnwidth,valign=t]{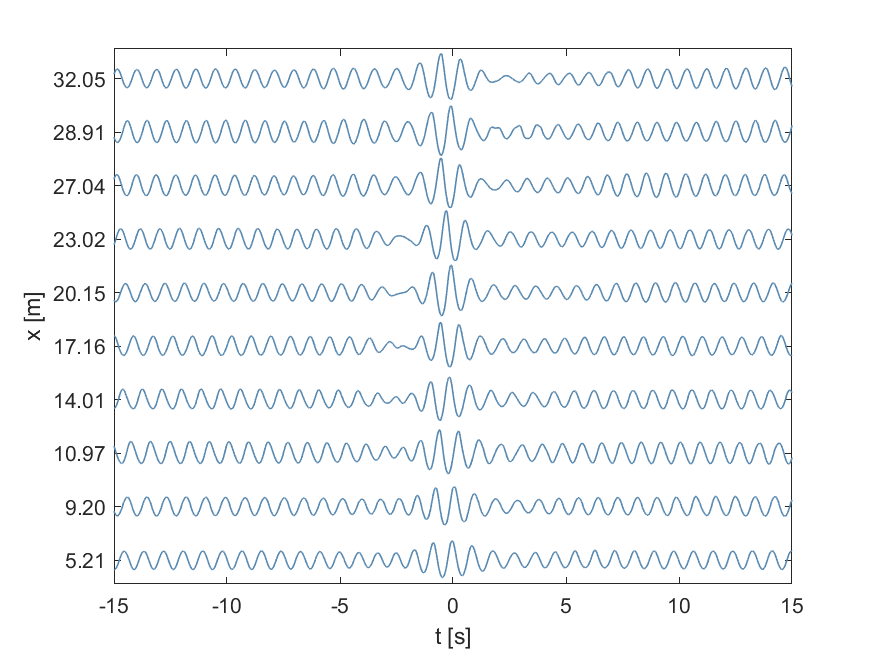}
\end{subfigure} \\
\begin{subfigure}[t]{0.015\columnwidth}\textbf{e}
\end{subfigure}
\begin{subfigure}[t]{0.4\columnwidth}
  \includegraphics[width=0.98\columnwidth,valign=t]{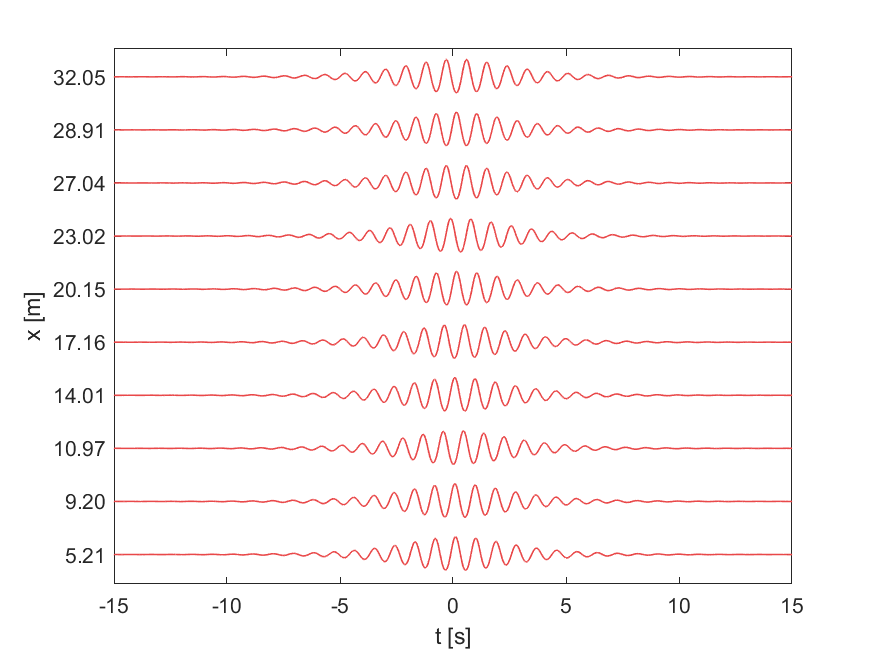} 
\end{subfigure}
\begin{subfigure}[t]{0.015\columnwidth}\textbf{f}
\end{subfigure}
\begin{subfigure}[t]{0.4\columnwidth}
	\includegraphics[width=0.98\columnwidth,valign=t]{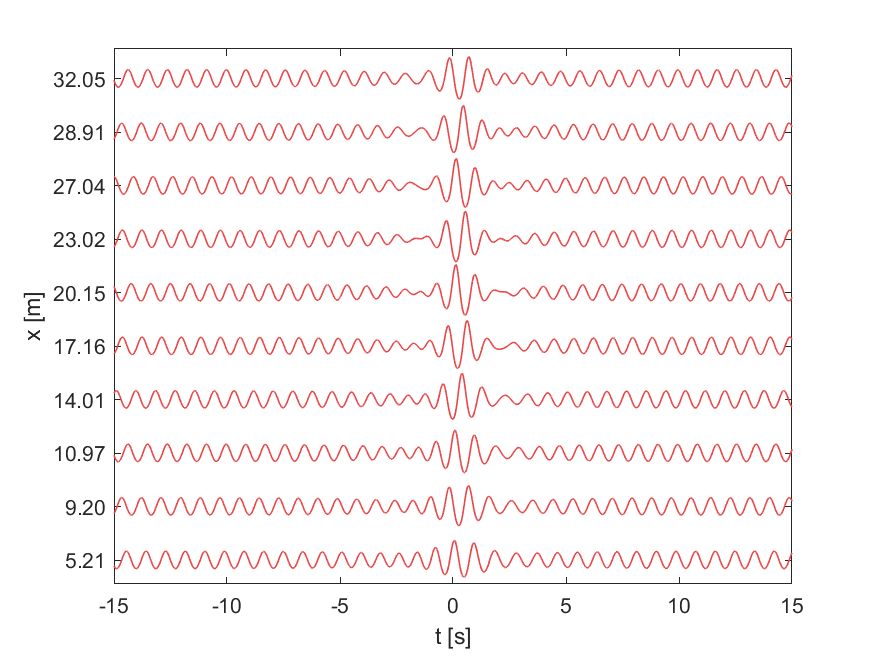}
\end{subfigure}
  \caption{Influence of the obliqueness angle on the temporal width of two localized structures for representative carrier parameters of $\boldmath a=0.02$ m and $\boldmath ak=0.1$. {\bf a} Theoretical envelope soliton surface for three different obliqueness angles. {\bf b} theoretical Peregrine breather surface for the sames angle values as in {\bf a}. {\bf c} Temporal water surface profiles of the envelope soliton for $\vartheta=20^\circ$ as measured in basin. {\bf d} Temporal water surface profiles of the Peregrine breather for $\vartheta=20^\circ$ as measured in basin. {\bf e} Theoretical water surface profiles corresponding to {\bf c}. {\bf f} Theoretical water surface profiles corresponding to {\bf d}.}
\label{fig4}
\end{figure} 

The measured evolution of a sech-type envelope soliton \cite{shabat1972exact} as well as the results obtained from the (2D+1) NLSE prediction are illustrated in Fig. \ref{fig3}. Each corresponding six plots show the oblique propagation of the localized and coherent structure with an angle of $\vartheta=20^\circ$ with respect to the carrier wave with parameters $ak=0.2$ and $a=0.02$ for the time interval of $\Delta t=0.39$ s starting at $t_0=67.04$ s, that is, $t_n=t_0+n\Delta t$, $n=0,\ldots,5$. Indeed, the direct comparison of the experimentally captured slanted envelope soliton in Fig. \ref{fig3}{\bf a} with the analytical (2D+1) NLSE prediction in Fig. \ref{fig3}{\bf b} reveals a very good agreement. This becomes particularly clear when comparing the accuracy of the phase as well as group velocities of the soliton propagation in the two cases, their amplitudes, and particularly the short crest lengths of the waves in each time frame.

The obliqueness angle $\vartheta$ also influences other parameters of the localized solutions. In particular, it affects the shape and the width of the soliton as well as the crest length of the directional wave field. Fig. \ref{fig4}{\bf a}. shows the case of the envelope soliton while Fig. \ref{fig4}{\bf b} the Peregrine breather envelope profiles as functions in time for several angles of propagation $\vartheta=0^\circ$, $\vartheta=20^\circ$ and $\vartheta=35^\circ$. 

The soliton becomes thinner with increasing the angle of propagation. The same applies to the Peregrine solution. Again, the profile of the solution compresses with increasing the angle. In case of periodic solutions, such as Akhmediev breathers or modulation instability in a general context, the period of the modulation will be also compressed.

In view of this angle-dependent compression, adjusted group velocity and due to the complexities in the marker-net evaluation of the data, we restricted ourselves to the wave gauge measurements along the flume and $x$-direction. 

Fig. \ref{fig4}{\bf c} displays the evolution of the slanted envelope soliton for $ak=0.1$,  $a=0.02$ m and $\vartheta=20^\circ$, while the panel in Fig. \ref{fig4}{\bf e} shows the corresponding curves calculated theoretically. 
The agreement between the experimental data and theoretical predictions is striking, especially, when considering the preservation of the coherence and taking into account relatively large propagation distance of the soliton. 

The oblique geometry also influences pulsating solutions localized in the propagation direction such as the Peregrine breather. Our equipment allowed us to generate them for a wide range of angles of propagation. Videos showing the evolution of the Peregrine and Akhmediev breathers can be found in the Supplementary Material. These videos clearly exemplify that the breather propagation direction differs from the carrier propagation direction just as in the case of the soliton. Difference of these directions is the major result of our observations. This discovery proves that localized, short-crested and directional water waves, particularly rogue waves, can be also described by a nonlinear framework.

The Peregrine solution can be considered as the limit of the Akhmediev breather, the analytical and deterministic modulation instability model, when the period of the modulation tends to infinity \cite{kibler2010peregrine,chabchoub2011rogue}. Then, maxima of the periodic modulated structure are well separated and only one localized peak remains at the center. The temporal evolution of a slanted Peregrine solution measured in the experiment is shown and compared to the (2D+1) NLSE predicted wave curves in Fig. \ref{fig4}{\bf d} and {\bf f}, respectively. Again, comparison of the corresponding panel pair in Fig. \ref{fig4} shows a remarkably good agreement between the measurements and the directional NLSE theory. The measured and calculated focusing distances, the maximal amplitudes and the width of this localized and pulsating solution are all in excellent agreement at all stage of propagation. 

\section*{Discussion}
Overall, our results reveal the existence of nonlinear solitary wave packets and breather beams, propagating obliquely to the direction of the wave field. This fact is confirmed by our experimental measurements for surface gravity water waves in a deep and directional water wave facility, installed at the University of Tokyo. Video files attached in the supplementary material clearly demonstrate and visualize this particular novel feature of nonlinear wave dynamics. The evolution of these packets is in excellent agreement with the (2D+1) NLSE framework in regard of all wave features. A remarkable property of these particular localized wave packets studied here is their finite crest length. The latter can be observed by simply watching the ocean waves. The crest length and thus, the transverse size of the waves is always limited. Now, it turns out that coherent waves with finite crest length might be a consequence of nonlinear beam dynamics. This is an important observation especially for the breather solutions, as this suggests that the nonlinearity is also a possible underlying mechanism for the actual finite-length-crested rogue wave events complementing the linear superposition and interference arguments as has been generally suggested. Further studies using a fully nonlinear hydrodynamic approach \cite{slunyaev2013super,sanina2016detection} may increase the accuracy of the description. These will characterize the ranges of accuracy of the approach, however, will not add anything substantial to the concept. Serious implications of such wave packets in oceanography is an important aspect of our results. This includes directional wave modeling, swell propagation and diffraction as well as remote sensing of waves to name few. Moreover, investigating wave breaking processes \cite{iafrati2013modulational,derakhti2018predicting} and prediction \cite{farazmand2017reduced,randoux2016inverse} of extreme directional waves is also crucial for future application purposes. Since the effect that can be explained by means of a general and universal theory for two-dimensional nonlinear wave fields in dispersive environments, its further extensions can stimulate analogous theoretical, numerical and experimental studies in two-dimensional optical surfaces, multi-dimensional plasmas, among other relevant physical media, elevating our level of understanding of these phenomena.

\section*{Methods} 
Our theoretical framework is based on the space-(2D+1) NLSE for deep-water waves \cite{zakharov1968stability}. For a wave envelope $\psi(x,y,t)$ with carrier wavenumber $k$ along the $x$-direction and carrier frequency $\omega=\sqrt{gk}$, we have
\begin{equation}
i\left(\frac{\partial \psi}{\partial t}+c_g\frac{\partial \psi}{\partial x}\right)
 -\lambda \frac{\partial^2 \psi}{\partial x^2} + 2 \lambda \frac{\partial^2 \psi}{\partial y^2}
-\gamma |\psi|^2 \psi=0 \label{NLS2D}
\end{equation}
where $\lambda=\frac{\omega}{8 k^2}$, $\gamma=\frac{\omega k^2}{2}$ and $g$ denotes the gravitational acceleration. At the leading order, it is known that $\frac{\partial \psi}{\partial t}\simeq c_g\frac{\partial \psi}{\partial x}$. This relation can be used to write the equation to express the wave packet propagation in space along the spatial $x$ co-ordinate to give a time-(2D+1) NLSE \cite{osborne2010nonlinear}
\begin{equation}
i\left(\frac{\partial\psi}{\partial x}+\frac{1}{c_g} \frac{\partial\psi}{\partial t}\right)
 -\frac{\lambda}{c_g^3} \frac{\partial^2 \psi}{\partial t^2} + 2 \frac{\lambda}{c_g} \frac{\partial^2 \psi}{\partial y^2}
-\frac{\gamma}{c_g} |\psi|^2 \psi=0 \label{NLS2Dtime}
\end{equation}
As the measurements are made at fixed positions along the flume, this equation can be used for experimental investigations. Now, we introduce the following transformation 
\begin{equation}
%T=(t-\frac{x}{c_g}) \cos\vartheta+ \frac{y}{c_g} \sin\vartheta
T=t\cos\vartheta-\frac{y}{c_g} \sin\vartheta \label{transformation}
\end{equation}
%{\color{red} there should be a + in the above formula??}
with variable parameter $\vartheta$ that sets a special relation between time, $t$, and the spatial coordinate $y$. Then, the 
evolution equation for the new wave function, $\psi(x, T)$, reads
%% TW: it might be easier to follow if a different symbol is used here for \psi
\begin{equation}
i\left(\frac{\partial \psi}{\partial x}+\frac{1}{C_g} \frac{\partial \psi}{\partial T}\right)
 -\Lambda\frac{\partial^2 \psi}{\partial T^2}-\Gamma |\psi|^2 \psi=0,\label{NLS2Dtimeslantedp}
\end{equation}
with $C_g=c_g/\cos \vartheta$, $\Lambda=\lambda(1-3 \sin^2\vartheta)/c_g^3$ and $\Gamma=\gamma/c_g$.
When the angle $|\vartheta|<\sqrt{\arcsin\left(1/3\right)}\simeq35.26^\circ$ \cite{saffman1978stability,yuen1982nonlinear},  Eq. (\ref{NLS2Dtimeslantedp}) is the standard (1D+1) focusing NLS equation that is known to be integrable \cite{shabat1972exact,akhmediev1997solitons,ablowitz2011nonlinear}.  %\cite{chabchoub2016hydrodynamic} 
When $\theta \ne 0$ the envelope and the phase travel at a finite angle to each other.

From an experimental point of view, the boundary condition for the surface elevation $\eta(x,y,t)$ at the wave maker, placed at $x=0$, can be described, to the leading-order, by the following expression:
\begin{equation}
\eta(x=0,y,t)=\frac{1}{2}\left[\psi(0,T) \exp\left({-i\omega t}\right)+c.c.\right],
\label{surface_el}
\end{equation}
where $\psi(0,T)$ is the desired solution of the one dimensional NLSE in Eq. (\ref{NLS2Dtimeslantedp}) and $T$ is given by Eq. (\ref{transformation}). Eq. (\ref{surface_el}) is used for driving the wave maker.

%\bibliography{Refs}
%\bibliographystyle{Science} 
%
%\section*{Captions for Movies}
%{\bf Movie S1}\\ 
%Visual representation and inspection of the propagation of oblique periodic Akhmediev breather rogue waves. This movie clearly shows that the breather direction differs from the carrier propagation orientation. The carrier parameters adopted are $a=0.01$ and $ak=0.1$ and the breather angle is $\vartheta=20^\circ$ while the modulation frequency has been chosen to satisfy the maximal instability growth rate.\\
%$
%$\\
%{\bf Movie S2}\\ 
%Visual representation and inspection of the propagation of an oblique doubly-localized Peregrine breather for the carrier parameters $a=0.02$ m and $ak=0.1$. This movie also exemplifies the diagonal breather beam dynamics as well as the short-crested modulated and large-amplitude waves located in the unstable wave packet.

%Being the first observation of the effect, our results may  in other nonlinear dispersive media, such as optics, plasma and BEC elevating our level of understanding of these phenomena.

\bibliography{Refs}

\begin{thebibliography}{10}
\expandafter\ifx\csname url\endcsname\relax
  \def\url#1{\texttt{#1}}\fi
\expandafter\ifx\csname urlprefix\endcsname\relax\def\urlprefix{URL }\fi
\providecommand{\bibinfo}[2]{#2}
\providecommand{\eprint}[2][]{\url{#2}}

\bibitem{komen1996dynamics}
\bibinfo{author}{Komen, G.~J.} \emph{et~al.}
\newblock \emph{\bibinfo{title}{Dynamics and modelling of ocean waves}}
  (\bibinfo{publisher}{Cambridge university press}, \bibinfo{year}{1996}).

\bibitem{rintoul2018global}
\bibinfo{author}{Rintoul, S.~R.}
\newblock \bibinfo{title}{The global influence of localized dynamics in the
  southern ocean.}
\newblock \emph{\bibinfo{journal}{Nature}} \textbf{\bibinfo{volume}{558}},
  \bibinfo{pages}{209--218} (\bibinfo{year}{2018}).

\bibitem{osborne1980internal}
\bibinfo{author}{Osborne, A.} \& \bibinfo{author}{Burch, T.}
\newblock \bibinfo{title}{Internal solitons in the andaman sea}.
\newblock \emph{\bibinfo{journal}{Science}} \textbf{\bibinfo{volume}{208}},
  \bibinfo{pages}{451--460} (\bibinfo{year}{1980}).

\bibitem{onorato2009statistical}
\bibinfo{author}{Onorato, M.} \emph{et~al.}
\newblock \bibinfo{title}{Statistical properties of directional ocean waves:
  the role of the modulational instability in the formation of extreme events}.
\newblock \emph{\bibinfo{journal}{Phys. Rev. Lett.}}
  \textbf{\bibinfo{volume}{102}}, \bibinfo{pages}{114502}
  (\bibinfo{year}{2009}).

\bibitem{toffoli2011extreme}
\bibinfo{author}{Toffoli, A.} \emph{et~al.}
\newblock \bibinfo{title}{Extreme waves in random crossing seas: Laboratory
  experiments and numerical simulations}.
\newblock \emph{\bibinfo{journal}{Geophysical Research Lett.}}
  \textbf{\bibinfo{volume}{38}}, \bibinfo{pages}{L06605}
  (\bibinfo{year}{2011}).

\bibitem{pinho2015emergence}
\bibinfo{author}{Pinho, U.~F.} \& \bibinfo{author}{Babanin, A.~V.}
\newblock \bibinfo{title}{Emergence of short crestedness in originally
  unidirectional nonlinear waves}.
\newblock \emph{\bibinfo{journal}{Geophysical Research Letters}}
  \textbf{\bibinfo{volume}{42}}, \bibinfo{pages}{4110--4115}
  (\bibinfo{year}{2015}).

\bibitem{zakharov1968stability}
\bibinfo{author}{Zakharov, V.~E.}
\newblock \bibinfo{title}{Stability of periodic waves of finite amplitude on
  the surface of a deep fluid}.
\newblock \emph{\bibinfo{journal}{J. Appl. Mech. Techn. Phys.}}
  \textbf{\bibinfo{volume}{9}}, \bibinfo{pages}{190--194}
  (\bibinfo{year}{1968}).

\bibitem{akhmediev1997solitons}
\bibinfo{author}{Akhmediev, N.} \& \bibinfo{author}{Ankiewicz, A.}
\newblock \emph{\bibinfo{title}{{Solitons: Nonlinear pulses and beams}}}
  (\bibinfo{publisher}{Chapman \& Hall, London}, \bibinfo{year}{1997}).

\bibitem{osborne2010nonlinear}
\bibinfo{author}{Osborne, A.}
\newblock \emph{\bibinfo{title}{Nonlinear Ocean Waves \& the Inverse Scattering
  Transform}}, vol.~\bibinfo{volume}{97} (\bibinfo{publisher}{Academic Press},
  \bibinfo{year}{2010}).

\bibitem{blanco2016pure}
\bibinfo{author}{Blanco-Redondo, A.} \emph{et~al.}
\newblock \bibinfo{title}{Pure-quartic solitons}.
\newblock \emph{\bibinfo{journal}{Nature communications}}
  \textbf{\bibinfo{volume}{7}}, \bibinfo{pages}{10427} (\bibinfo{year}{2016}).

\bibitem{solli2007optical}
\bibinfo{author}{Solli, D.}, \bibinfo{author}{Ropers, C.},
  \bibinfo{author}{Koonath, P.} \& \bibinfo{author}{Jalali, B.}
\newblock \bibinfo{title}{Optical rogue waves}.
\newblock \emph{\bibinfo{journal}{Nature}} \textbf{\bibinfo{volume}{450}},
  \bibinfo{pages}{1054--1057} (\bibinfo{year}{2007}).

\bibitem{onorato2013roguereport}
\bibinfo{author}{Onorato, M.}, \bibinfo{author}{Residori, S.},
  \bibinfo{author}{Bortolozzo, U.}, \bibinfo{author}{Montina, A.} \&
  \bibinfo{author}{Arecchi, F.~T.}
\newblock \bibinfo{title}{Rogue waves and their generating mechanisms in
  different physical contexts}.
\newblock \emph{\bibinfo{journal}{Physics Reports}}
  \textbf{\bibinfo{volume}{528}}, \bibinfo{pages}{47--89}
  (\bibinfo{year}{2013}).

\bibitem{dudley2014instabilities}
\bibinfo{author}{Dudley, J.~M.}, \bibinfo{author}{Dias, F.},
  \bibinfo{author}{Erkintalo, M.} \& \bibinfo{author}{Genty, G.}
\newblock \bibinfo{title}{Instabilities, breathers and rogue waves in optics}.
\newblock \emph{\bibinfo{journal}{Nature Photonics}}
  \textbf{\bibinfo{volume}{8}}, \bibinfo{pages}{755--764}
  (\bibinfo{year}{2014}).

\bibitem{kartashov2011solitons}
\bibinfo{author}{Kartashov, Y.~V.}, \bibinfo{author}{Malomed, B.~A.} \&
  \bibinfo{author}{Torner, L.}
\newblock \bibinfo{title}{Solitons in nonlinear lattices}.
\newblock \emph{\bibinfo{journal}{Reviews of Modern Physics}}
  \textbf{\bibinfo{volume}{83}}, \bibinfo{pages}{247} (\bibinfo{year}{2011}).

\bibitem{saffman1978stability}
\bibinfo{author}{Saffman, P.} \& \bibinfo{author}{Yuen, H.~C.}
\newblock \bibinfo{title}{Stability of a plane soliton to infinitesimal
  two-dimensional perturbations}.
\newblock \emph{\bibinfo{journal}{Phys. Fluids}} \textbf{\bibinfo{volume}{21}},
  \bibinfo{pages}{1450--1451} (\bibinfo{year}{1978}).

\bibitem{yuen1982nonlinear}
\bibinfo{author}{Yuen, H.~C.} \& \bibinfo{author}{Lake, B.~M.}
\newblock \bibinfo{title}{Nonlinear dynamics of deep-water gravity waves}.
\newblock \emph{\bibinfo{journal}{Adv. Appl. Mech}}
  \textbf{\bibinfo{volume}{22}}, \bibinfo{pages}{229} (\bibinfo{year}{1982}).

\bibitem{longuet1976nonlinear}
\bibinfo{author}{Longuet-Higgins, M.~S.}
\newblock \bibinfo{title}{On the nonlinear transfer of energy in the peak of a
  gravity-wave spectrum: a simplified model}.
\newblock \emph{\bibinfo{journal}{Proc. R. Soc. Lond. A}}
  \textbf{\bibinfo{volume}{347}}, \bibinfo{pages}{311--328}
  (\bibinfo{year}{1976}).

\bibitem{toffoli2010development}
\bibinfo{author}{Toffoli, A.}, \bibinfo{author}{Onorato, M.},
  \bibinfo{author}{Bitner-Gregersen, E.} \& \bibinfo{author}{Monbaliu, J.}
\newblock \bibinfo{title}{Development of a bimodal structure in ocean wave
  spectra}.
\newblock \emph{\bibinfo{journal}{J. Geophysical Research: Oceans}}
  \textbf{\bibinfo{volume}{115}}, \bibinfo{pages}{C03006}
  (\bibinfo{year}{2010}).

\bibitem{borge2013detection}
\bibinfo{author}{Borge, J.~N.}, \bibinfo{author}{Reichert, K.} \&
  \bibinfo{author}{Hessner, K.}
\newblock \bibinfo{title}{Detection of spatio-temporal wave grouping properties
  by using temporal sequences of x-band radar images of the sea surface}.
\newblock \emph{\bibinfo{journal}{Ocean Modelling}}
  \textbf{\bibinfo{volume}{61}}, \bibinfo{pages}{21--37}
  (\bibinfo{year}{2013}).

\bibitem{nguyen2017formation}
\bibinfo{author}{Nguyen, J.~H.}, \bibinfo{author}{Luo, D.} \&
  \bibinfo{author}{Hulet, R.~G.}
\newblock \bibinfo{title}{Formation of matter-wave soliton trains by
  modulational instability}.
\newblock \emph{\bibinfo{journal}{Science}} \textbf{\bibinfo{volume}{356}},
  \bibinfo{pages}{422--426} (\bibinfo{year}{2017}).

\bibitem{cabrera2018quantum}
\bibinfo{author}{Cabrera, C.} \emph{et~al.}
\newblock \bibinfo{title}{Quantum liquid droplets in a mixture of bose-einstein
  condensates}.
\newblock \emph{\bibinfo{journal}{Science}} \textbf{\bibinfo{volume}{359}},
  \bibinfo{pages}{301--304} (\bibinfo{year}{2018}).

\bibitem{cheiney2018bright}
\bibinfo{author}{Cheiney, P.} \emph{et~al.}
\newblock \bibinfo{title}{Bright soliton to quantum droplet transition in a
  mixture of bose-einstein condensates}.
\newblock \emph{\bibinfo{journal}{Physical review letters}}
  \textbf{\bibinfo{volume}{120}}, \bibinfo{pages}{135301}
  (\bibinfo{year}{2018}).

\bibitem{semeghini2018self}
\bibinfo{author}{Semeghini, G.} \emph{et~al.}
\newblock \bibinfo{title}{Self-bound quantum droplets of atomic mixtures in
  free space}.
\newblock \emph{\bibinfo{journal}{Physical Review Letters}}
  \textbf{\bibinfo{volume}{120}}, \bibinfo{pages}{235301}
  (\bibinfo{year}{2018}).

\bibitem{kartashov2018three}
\bibinfo{author}{Kartashov, Y.~V.}, \bibinfo{author}{Malomed, B.~A.},
  \bibinfo{author}{Tarruell, L.} \& \bibinfo{author}{Torner, L.}
\newblock \bibinfo{title}{Three-dimensional droplets of swirling superfluids}.
\newblock \emph{\bibinfo{journal}{Physical Review A}}
  \textbf{\bibinfo{volume}{98}}, \bibinfo{pages}{013612}
  (\bibinfo{year}{2018}).

\bibitem{shabat1972exact}
\bibinfo{author}{Shabat, A.~B.} \& \bibinfo{author}{Zakharov, V.~E.}
\newblock \bibinfo{title}{Exact theory of two-dimensional self-focusing and
  one-dimensional self-modulation of waves in nonlinear media}.
\newblock \emph{\bibinfo{journal}{Sov. Phys. JETP}}
  \textbf{\bibinfo{volume}{34}}, \bibinfo{pages}{62} (\bibinfo{year}{1972}).

\bibitem{peregrine1983water}
\bibinfo{author}{Peregrine, D.~H.}
\newblock \bibinfo{title}{{Water waves, nonlinear Schr{\"o}dinger equations and
  their solutions}}.
\newblock \emph{\bibinfo{journal}{J. Australian Math. Soc. Series B. Applied
  Mathematics}} \textbf{\bibinfo{volume}{25}}, \bibinfo{pages}{16--43}
  (\bibinfo{year}{1983}).

\bibitem{mozumi20153d}
\bibinfo{author}{Mozumi, K.}, \bibinfo{author}{Waseda, T.} \&
  \bibinfo{author}{Chabchoub, A.}
\newblock \bibinfo{title}{3d stereo imaging of abnormal waves in a wave basin}.
\newblock In \emph{\bibinfo{booktitle}{ASME 2015 34th International Conference
  on Ocean, Offshore and Arctic Engineering}},
  \bibinfo{pages}{V003T02A027--V003T02A027} (\bibinfo{organization}{American
  Society of Mechanical Engineers}, \bibinfo{year}{2015}).

\bibitem{kibler2010peregrine}
\bibinfo{author}{Kibler, B.} \emph{et~al.}
\newblock \bibinfo{title}{{The Peregrine soliton in nonlinear fibre optics}}.
\newblock \emph{\bibinfo{journal}{Nature Physics}}
  \textbf{\bibinfo{volume}{6}}, \bibinfo{pages}{790--795}
  (\bibinfo{year}{2010}).

\bibitem{chabchoub2011rogue}
\bibinfo{author}{Chabchoub, A.}, \bibinfo{author}{Hoffmann, N.} \&
  \bibinfo{author}{Akhmediev, N.}
\newblock \bibinfo{title}{Rogue wave observation in a water wave tank}.
\newblock \emph{\bibinfo{journal}{Phys. Rev. Lett.}}
  \textbf{\bibinfo{volume}{106}}, \bibinfo{pages}{204502}
  (\bibinfo{year}{2011}).

\bibitem{slunyaev2013super}
\bibinfo{author}{Slunyaev, A.} \emph{et~al.}
\newblock \bibinfo{title}{Super-rogue waves in simulations based on weakly
  nonlinear and fully nonlinear hydrodynamic equations}.
\newblock \emph{\bibinfo{journal}{Phys. Rev. E}} \textbf{\bibinfo{volume}{88}},
  \bibinfo{pages}{012909} (\bibinfo{year}{2013}).

\bibitem{sanina2016detection}
\bibinfo{author}{Sanina, E.}, \bibinfo{author}{Suslov, S.},
  \bibinfo{author}{Chalikov, D.} \& \bibinfo{author}{Babanin, A.}
\newblock \bibinfo{title}{Detection and analysis of coherent groups in
  three-dimensional fully-nonlinear potential wave fields}.
\newblock \emph{\bibinfo{journal}{Ocean Modelling}}
  \textbf{\bibinfo{volume}{103}}, \bibinfo{pages}{73--86}
  (\bibinfo{year}{2016}).

\bibitem{iafrati2013modulational}
\bibinfo{author}{Iafrati, A.}, \bibinfo{author}{Babanin, A.} \&
  \bibinfo{author}{Onorato, M.}
\newblock \bibinfo{title}{Modulational instability, wave breaking, and
  formation of large-scale dipoles in the atmosphere}.
\newblock \emph{\bibinfo{journal}{Phys. Rev. Lett.}}
  \textbf{\bibinfo{volume}{110}}, \bibinfo{pages}{184504}
  (\bibinfo{year}{2013}).

\bibitem{derakhti2018predicting}
\bibinfo{author}{Derakhti, M.}, \bibinfo{author}{Banner, M.~L.} \&
  \bibinfo{author}{Kirby, J.~T.}
\newblock \bibinfo{title}{Predicting the breaking strength of gravity water
  waves}.
\newblock \emph{\bibinfo{journal}{J. Fluid Mechanics}}
  \textbf{\bibinfo{volume}{848}}, \bibinfo{pages}{R2} (\bibinfo{year}{2018}).

\bibitem{farazmand2017reduced}
\bibinfo{author}{Farazmand, M.} \& \bibinfo{author}{Sapsis, T.~P.}
\newblock \bibinfo{title}{Reduced-order prediction of rogue waves in
  two-dimensional deep-water waves}.
\newblock \emph{\bibinfo{journal}{J. Computational Phys.}}
  \textbf{\bibinfo{volume}{340}}, \bibinfo{pages}{418--434}
  (\bibinfo{year}{2017}).

\bibitem{randoux2016inverse}
\bibinfo{author}{Randoux, S.}, \bibinfo{author}{Suret, P.} \&
  \bibinfo{author}{El, G.}
\newblock \bibinfo{title}{Inverse scattering transform analysis of rogue waves
  using local periodization procedure}.
\newblock \emph{\bibinfo{journal}{Scientific reports}}
  \textbf{\bibinfo{volume}{6}}, \bibinfo{pages}{29238} (\bibinfo{year}{2016}).

\bibitem{ablowitz2011nonlinear}
\bibinfo{author}{Ablowitz, M.~J.}
\newblock \emph{\bibinfo{title}{Nonlinear Dispersive Waves: Asymptotic Analysis
  and Solitons}}, vol.~\bibinfo{volume}{47} (\bibinfo{publisher}{Cambridge
  University Press}, \bibinfo{year}{2011}).

\end{thebibliography}
\bibliographystyle{naturemag} 
\section*{Acknowledgments}
A.C. acknowledges support from the Japan Society for the Promotion of Science (JSPS). N.A. acknowledges the Australian Research Council for financial support. M.O. has been funded by Progetto di Ricerca d'Ateneo CSTO160004. M.O. was supported by the ``Departments of Excellence 2018-2022'' Grant awarded by the Italian Ministry of Education, University and Research (MIUR) (L.232/2016). The experiment at the University of Tokyo was supported by Kakenhi of JSPS.

%\section*{Supplementary materials}
%Methods\\
%References \textit{(32-38)}\\
%Captions for Movies S1 to S2
\end{document}